\documentclass{article}
\usepackage{spconf,amsmath,graphicx}

\usepackage{color}
\usepackage{xcolor}
\usepackage{enumitem}
\usepackage{hyperref}
\usepackage[capitalize]{cleveref}
\usepackage{multirow}
\usepackage{etoolbox,siunitx}
\robustify\bfseries
\usepackage{booktabs}
\usepackage{balance}
\usepackage{amssymb}
\usepackage{bm}
\usepackage{arydshln}
\usepackage{caption}
\usepackage{subcaption}
\usepackage{cite}
\usepackage{hyphenat}

\DeclareMathOperator*{\argmin}{arg\,min}

\usepackage{soul}

\usepackage[acronym,shortcuts]{glossaries}
\glsdisablehyper    %

\newacronym{ASR}{ASR}{Automatic Speech Recognition}
\newacronym{WER}{WER}{Word Error Rate}
\newacronym{NN}{NN}{Neural Network}
\newacronym{RIR}{RIR}{Room Impulse Response}
\newacronym{STFT}{STFT}{Short-Time Fourier Transform}
\newacronym{MTF}{MTF}{Multiplication Transfer Function}
\newacronym{MSE}{MSE}{Mean Squared Error}

\title{Reverberation as supervision for speech separation}
\name{Rohith Aralikatti$^{1,2}$\!, Christoph Boeddeker$^{1,3}$\!, Gordon Wichern$^1$\!, Aswin Subramanian$^1$\!, Jonathan Le Roux$^1$\thanks{This work was performed while R.~Aralikatti and C.~Boeddeker were interns at MERL.}}
\address{$^1$Mitsubishi Electric Research Laboratories (MERL), Cambridge, MA, USA\\
$^{2}$University of Maryland, College Park, MD, USA \quad
$^{3}$Paderborn University, Paderborn, Germany
}
\begin{document}
\ninept
\maketitle
\begin{abstract}
This paper proposes reverberation as supervision (RAS), a novel unsupervised loss function for single-channel reverberant speech separation. Prior methods for unsupervised separation required the synthesis of mixtures of mixtures or assumed the existence of a teacher model, making them difficult to consider as potential methods explaining the emergence of separation abilities in an animal's auditory system. We assume the availability of two-channel mixtures at training time, and train a neural network to separate the sources given one of the channels as input such that the other channel may be predicted from the separated sources. As the relationship between the room impulse responses (RIRs) of each channel depends on the locations of the sources, which are unknown to the network, the network cannot rely on learning that relationship. Instead, our proposed loss function fits each of the separated sources to the mixture in the target channel via Wiener filtering, and compares the resulting mixture to the ground-truth one. We show that minimizing the scale-invariant signal-to-distortion ratio (SI-SDR) of the predicted right-channel mixture with respect to the ground truth implicitly guides the network towards separating the left-channel sources. On a semi-supervised reverberant speech separation task based on the WHAMR! dataset, using training data where just 5\% (resp., 10\%) of the mixtures are labeled with associated isolated sources, we achieve  70\% (resp., 78\%) of the SI-SDR improvement obtained when training with supervision on the full training set, while a model trained only on the labeled data obtains 43\% (resp., 45\%).
\end{abstract}
\section{Introduction}
\label{sec:intro}

The approaches of deep clustering~\cite{Hershey2016} and permutation-invariant training~\cite{kolbaek2017uPIT, Isik2016Interspeech09} facilitated an explosion of interest in learning to separate overlapped speech signals, a research field commonly known as speech separation. Additional advances included time domain models~\cite{luo2019convTasNet, heitkaemper2020demystifying, ditter2020multi} and powerful multi-channel systems~\cite{wang2018multi,wang2018integrating, gu2020enhancing}. Speech separation is now a standard front-end supporting speaker diarization~\cite{kinoshita2020tackling, Chen2020LibriCSS, Raj2021Meeting} and automatic speech recognition~\cite{settle2018end, haeb2020far} amongst other applications. However, full emulation of the human ability to solve the cocktail party problem, i.e., effortlessly focus on target speech in a noisy, multi-talker environment remains far off.

Part of the reason speech separation approaches remain far off from human-level performance is the reliance on fully-supervised approaches requiring vast quantities of labeled mixture data, which is typically generated synthetically. It is challenging to record both the mixture and the ground-truth speech signals in a real-world environment as it is difficult to suppress cross-talk and background noise. However, synthetically generated mixtures cannot accurately model many real-world phenomena such as reverberation and diffraction, material properties of acoustic environments, moving sources and listeners, etc. This domain mismatch between synthetic data and the actual data distribution often causes a reduction in performance. Hence, there is great value in formulating techniques that can leverage recorded, unlabeled mixtures to address this data gap.

\begin{figure}[t]
    \centering
        \includegraphics[width=1.0\linewidth]{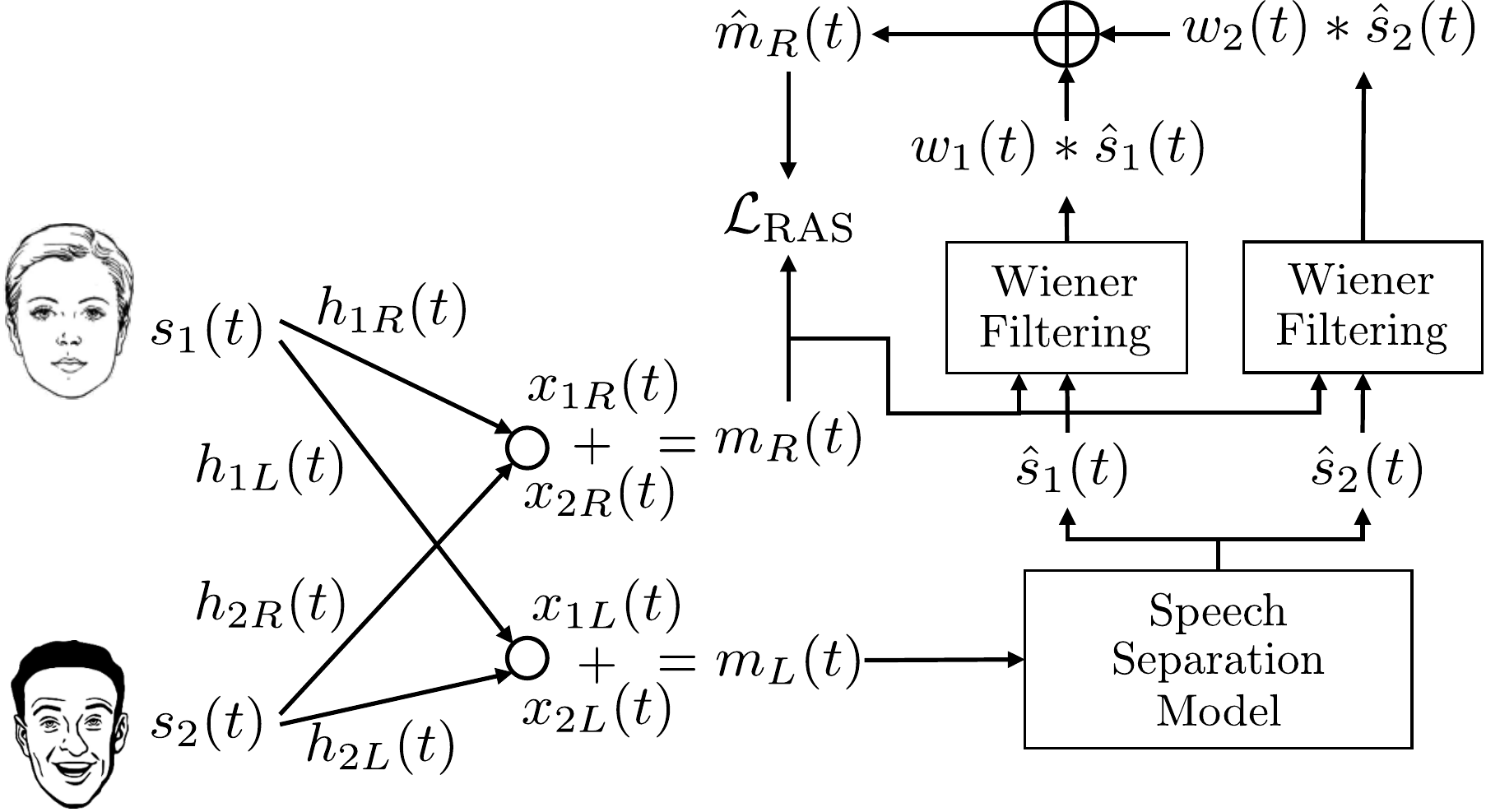}
    \caption{Illustration of the proposed reverberation as supervision approach.}\vspace{-.1cm}
    \label{fig:overall_example}
\end{figure}

Recently,  MixIT~\cite{mixit} demonstrated great success in unsupervised speech separation by creating ``mixtures of mixtures,’’ i.e., summing together multiple overlapped speech signals, separating them, and then using a permutation matrix to reconstruct the original mixtures which are compared to the input mixtures as a training objective. Further extensions of MixIT for real-world meeting data~\cite{sivaraman2022adapting} and RemixIT~\cite{tzinis2022remixit} for speech enhancement have confirmed its value for speech separation. However, despite the successes, there is something unsatisfying about MixIT from a biological perspective, i.e., it is hard to imagine the human brain learning to separate sounds by creating mixtures of mixtures. Humans use a variety of cues - contextual and perceptual cues, speaker lip movements, and spatial information to extract the required speech. In this work, we explore an approach where difference in the observed reverberation between two microphones (analogous to two ears) can be used as a supervision signal for speech signals as illustrated in Figure~\ref{fig:overall_example}.

Prior approaches for unsupervised speech separation based on spatial cues~\cite{drude2019unsupervised_icassp,seetharaman2019bootstrapping,tzinis2019unsupervised} use features such as inter-channel phase difference computed from two channel speech mixtures to create pseudo labels for training supervised speech separation models. Similar to the discussion on MixIT above, the existence of a teacher model is difficult to consider as a potential method explaining the emergence of separation abilities in an animal’s auditory system. The approach in~\cite{drude2019unsupervised_interspeech} does not make use of a teacher-student setup, but uses a large number of channels to learn a mask-based beamformer. Inspired by~\cite{arteaga2021multichannel}, which uses the strong inductive biases of an audio rendering system to train a multi-channel separation network in an unsupervised manner, we here propose reverberation as supervision (RAS) to train a neural network to separate speech sources given one channel in a two-channel mixture such that the other channel may be predicted from the separated sources. By fitting each of the separated sources to the mixture in the opposite channel via differentiable Wiener filtering~\cite{boeddeker2021convolutive}, we use the difference between the reconstructed and ground-truth opposite channel as our loss function. 

Through experiments on the WHAMR! dataset~\cite{maciejewski2020whamr} of two-channel reverberant two-speaker mixtures, we demonstrate the value of RAS in a semi-supervised setting. We also enumerate several design considerations necessary to obtain strong performance, such as the use of independent Wiener filters for each speaker, and filtering samples whose geometry is uninformative for the RAS objective. While we consider the two-channel two-source case for simplicity, the method could be readily extended to more channels (either to train a single-channel system, or a $K$ channel system from a dataset of $M$ channel mixtures, for $K<M$) and more sources.

\section{Reverberation as supervision (RAS)}

\subsection{Motivation}
We consider two dry source speech signals $s_1(t)$ and $s_2(t)$ originating from two speakers in different locations in a room, and a listener consisting of a two-microphone array at another location. For clarity of presentation and to stress the analogy with a binaural biological system, we refer to the individual channels as Left (L) and Right (R). The room impulse response from speaker $k\in\{1,2\}$ to microphone $c \in \{L,R\}$ is denoted as $h_{kc}(t)$. The source-image signal at each microphone is obtained as
\begin{equation}
x_{kc}(t) = h_{kc}(t) * s_k(t),
\end{equation}
where $*$ denotes convolution, and the observed mixture as
\begin{equation}
m_c(t) = \sum_k x_{kc}(t).
\end{equation}
The direct-path (anechoic) signal corresponding to $x_{kc}(t)$ is denoted as $d_{kc}(t)$.
For simplicity of presentation, we will consider the left channel as the input to the system, and the right channel as the supervision. At training time, we can alternately reverse their roles.

We are interested in deriving a loss function that pushes a neural network to separate the input mixture in the left channel into its constituent sources, by utilizing as an implicit measure of separation performance how well the observed mixture in the right channel can be predicted from the separated outputs. 
Because the network has no knowledge of the location of the sources, this prediction requires a relative impulse response to be applied in order to match the right channel. 
One way to do so is to apply a Wiener filter to the estimates.

In order to assess the viability of using the predicted fit to the right channel as an implicit measure of separation of the left channel, we first consider oracle experiments in which we fit the left-channel signals of the original input mixture as well as various versions of the ground-truth sources to the right-channel mixture via a Wiener filter, and evaluate the prediction performance. When fitting the two sources, the Wiener filter for each source is computed independently, and the filtered sources are added to form the mixture estimate $\hat{m}_R$. We use a Wiener filter with \num{512} coefficients, out of which \num{100} are non-causal. Non-causal filter weights are required as the right-channel signal may have a negative delay with respect to the left-channel signal if a source is closer to the right channel.
Table~\ref{table:oracle} shows the signal-to-distortion ratio (SDR) of the right-channel mixture estimate $\hat{m}_R$ with respect to the ground truth $m_R$ over the WHAMR! validation set. We see that there is an SDR improvement of \num{5} to \SI{6}{\decibel} when estimating the right-channel mixture from some version of the ground-truth left-channel sources as compared to estimating it from the left-channel mixture $m_L$ itself. We believe this difference is sufficient to push the neural network towards separating the sources when we optimize the speech separation network through the Wiener filter to reduce the SDR of the predicted right-channel mixture.  We also see that the direct-path signals lead to better reconstruction than the reverberant source-image signals, indicating that our objective may also incentivize some amount of dereverberation.  

\begin{table}[t]
\centering
    \sisetup{
    detect-weight, %
    mode=text, %
    tight-spacing=true,
    round-mode=places,
    round-precision=1,
    table-format=2.1,
    table-text-alignment=right
    }
\caption{\label{tab:oracle} Oracle SDR [dB] when predicting the right channel mixture from various left channel signals using a Wiener filter. %
We see that the filter models the output mixture more accurately when separated speech signals are given as input. Predicting the right channel mixture from the left channel mixture leads to a reduction of roughly \SI[round-precision=0]{5}{\decibel} in SDR.}\label{table:oracle}
\vspace{-.3cm}
\begin{tabular}{lS}
\toprule
Wiener filter input & \multicolumn{1}{r}{SDR($m_R$, $\hat{m}_R$)}   \\
\midrule
Mixture $m_L$        &  6.9   \\
Reverberant source-image signals $x_{1L}$, $x_{2L}$                 & 12.0 \\
Dry source signals $s_{1}$, $s_{2}$                                    & 13.2 \\
Direct-path signals $d_{1L}$, $d_{2L}$                                  & 13.9  \\
\bottomrule
\end{tabular}
\vspace{-.3cm}
\end{table}

\subsection{Filter design considerations}\label{ssec:filter}

As the network has no knowledge of the relative \gls{RIR} of each source from the left channel to the right channel, we need to allow some filtering on the separated signals to fit the right-channel mixture. We need to carefully design the filter in order to allow the right amount of flexibility, as too much flexibility would allow any signal to fit the right-channel mixture, while too little flexibility would prevent any meaningful fit, which in both cases would result in an uninformative objective. Design choices to consider include in particular the filter length and the split between causal and non-causal coefficients, and the way the filter is optimized. The filters we consider are FIR filters.

\noindent {\bf Causality and filter length:}
The longer the filter, the better it can fit a target signal with less dependence on the signal being filtered. We are not concerned with perfectly reproducing the relative RIR between the left and right channels, and focus on reproducing the effect of its early parts. As a source may be closer to the right channel than the left one, we also need to allow for non-causal coefficients so that the filtered signal may start earlier than the signal being filtered. 
We thus assume that a linear filter $w_{k}(t)$ lies between a good estimate for the $k$th speaker $\hat{x}_{kL}(t)$ of the left microphone and the signal of the same speaker at the other microphone $x_{kR}(t)$:
\begin{align}
    \hat{x}_{kR}(t) = (w_k * \hat{x}_{kL})(t) = \sum_{\tau = -\tau_{\text{nc}}}^{\tau_{\text{c}}-1} w_k(\tau) \hat{x}_{kL}(t - \tau) ,
\end{align}
where $\tau_{\text{nc}}$ denotes the number of non-causal coefficients, and $\tau_{\text{c}}$ the number of causal coefficients. In practice, we set $\tau_{\text{nc}}=100$ and $\tau_{\text{c}}=412$.

\noindent {\bf Joint vs independent filter estimation:}
To estimate the filter $w_{k}(t)$, a common objective is the sum of the squared errors
\begin{align}
    \argmin_{w_k(t)} \sum_t | (w_k * \hat{x}_{kL})(t) - x_{kR}(t) |^2,
\end{align}
which yields the solution of the Wiener-Hopf equation.

This estimation requires supervision, but when we jointly estimate the filters for each speaker (i.e., we sum the estimates), we could use the observation $m_{R}(t)$ as target in the estimation:
\begin{align}
    \argmin_{w_1(t), w_2(t)} \sum_t \left| \sum_k (w_k * \hat{x}_{kL})(t) - m_{R}(t)  \right|^2.
\end{align}
This equation is well suited to estimate the filters $w_k(t)$, but has issues with the gradients for $\hat{x}_{kL}(t)$, which are necessary for the \gls{NN} training. Indeed, any linear combination of the estimates $\hat{x}_{kL}(t)$ will yield the same 
score of the objective,
if the linear equation system has a full rank (i.e., the transformation matrix is invertible).

To partially address this issue, we consider an independent estimation:
\begin{align}
    \argmin_{w_k(t)} \sum_t | (w_k * \hat{x}_{kL})(t) - m_{R}(t)  |^2, \forall k, \label{eq:indep_Wiener}
\end{align}
while keeping the estimation of $\hat{m}_{R}(t)$ as:
\begin{align}
    \hat{m}_{R}(t) = w_1(t) * \hat{x}_{1L}(t) + w_2(t) * \hat{x}_{2L}(t). \label{eq:mR_rec}
\end{align}
It can be shown that
this independent estimation promotes some notion of uncorrelatedness between the estimates, which is a stronger constraint than with the joint estimation.
It however does not by itself fully constrain the network to output statistically independent estimates $\hat{x}_{kL}$ that are close to each source-image signal, which is why we will assume the availability of a small amount of supervised data in order to nudge the network towards the proper solution.
More advanced techniques for obtaining statistically independent estimates, such as independent component analysis (ICA) \cite{Hyvarinen2002ICA}, may be considered in future works to introduce a stronger inductive bias and avoid this issue.

\vspace{-.1cm}
\subsection{RAS objective}
\label{ssec:rasObjective}
\vspace{-.05cm}

Our goal is to perform single-channel speech separation to estimate the separated signals $x_{1L}(t)$ and $x_{2L}(t)$ given an observed mixture at the left channel $m_L(t)$. 
In our proposed method, we first obtain estimates $\hat{x}_{kL}(t)$ from a neural network given $m_L(t)$ as input. Wiener filtering is used to independently fit each $\hat{x}_{kL}(t)$ to the right-channel mixture ${m}_R(t)$ as in Eq.~(\ref{eq:indep_Wiener}), using the filtered signals to obtain an estimate $\hat{m}_R(t)$ of the right-channel mixture as in Eq.~(\ref{eq:mR_rec}). 
The Reverberation-as-supervision (RAS) loss is then defined as:
\begin{equation}
    \label{eq:ras}
    \mathcal{L}_{\operatorname{RAS}}(m_R,\hat{x}_{1L},\hat{x}_{2L}) = \mathcal{L}_{\operatorname{SI-SDR}}(m_R(t), \hat{m}_R(t)),
\end{equation}
where $\mathcal{L}_{\operatorname{SI-SDR}}$ is the SI-SDR loss \cite{leroux2019sdr,luo2019convTasNet,vincent2006bss} defined as
\begin{equation}
    \mathcal{L}_{\operatorname{SI-SDR}}(x, \hat{x}) = - 10 \log_{10}\frac{\|\alpha x\|^2}{\|\alpha x - \hat{x}\|^2}, \; \alpha = \frac{\langle x, \hat{x}\rangle}{\|x\|^2}.
\end{equation}
As no information about the ground-truth isolated signals is used in $\mathcal{L}_{RAS}$, it can be applied to mixtures for which no ground-truth signals are available. 
Assuming the availability of a small amount of labeled data (single-channel mixtures with corresponding isolated signals), we can combine $\mathcal{L}_{\operatorname{SI-SDR}}$ on the labeled data with $\mathcal{L}_{\operatorname{RAS}}$ on unlabeled data (two-channel mixtures) to obtain a semi-supervised approach for training speech separation models.
In our experiments, we show that this semi-supervised approach leads to impressive results when compared to a fully-supervised setting with just a fraction (\SI{5}{\percent}~-~\SI{10}{\percent}) of supervised data.

\vspace{-.1cm}
\subsection{Data selection to remove uninformative examples}
\vspace{-.05cm}

While applying RAS, we find it beneficial to filter out mixtures that had a very high correlation between the left and right channel mixtures. Intuitively, if it is easy to predict the right-channel mixture from the left-channel mixture, then there is not much left to gain for the network by separating the sources to further improve the right-channel prediction. These examples may consist for example of situations in which the sources are located in similar positions with respect to the two channels.

\begin{figure}[t]
    \centering
    \includegraphics[width=\columnwidth]{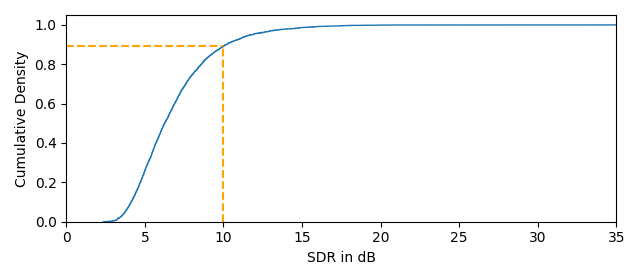}
    \vspace{-.7cm}
    \caption{Cumulative density function (CDF) of $\mathrm{SDR}(m_R(t), \hat{m}_R(t))$ with $\hat{m}_R(t)$ estimated from the left channel mixture $m_L(t)$ using a Wiener filter of size \num{512} with \num{100} non-causal weights and $m_R(t)$ as the reference signal. We discard mixtures for which $\text{SDR}(m_R(t), \hat{m}_R(t))$ is above \SI{10}{dB} (roughly \SI{10}{\percent} of the mixtures in the training data) during fine-tuning, as the left and right channels are too similar.}\vspace{-.4cm}
    \label{fig:SDR_histogram}
\end{figure}

We thus first estimate the right-channel mixture from the left-channel mixture using a Wiener filter, and compute the SDR between the estimate and the observed right-channel.
We filter out mixtures by thresholding this SDR value.
Empirically, we found that using mixtures that had an SDR lower than \SI{10}{dB} for fine-tuning gave the best results. The cumulative density function (CDF) of SDR for mixtures present in the validation set is shown in Fig.~\ref{fig:SDR_histogram}.

\vspace{-.1cm}
\section{Experimental validation}
\subsection{Dataset and experimental setup}
All experiments in this paper are done on the dual-channel (microphone spacing between \num{15}~-~\SI{17}{cm}), noise-free, \SI{8}{kHz} min version of the WHAMR! \cite{maciejewski2020whamr} speech separation dataset. Realistic speech mixtures in WHAMR! are generated by accurately simulating room reverberation effects and by adding background noise recorded in a variety of environments. Here, we do not add noise, and attempt to separate a reverberant mixture into its constituent reverberant sources. Our proposed approach assumes that a convolutive transfer function maps signals from one channel to the other. The addition of noise would weaken that assumption, and is left to future work.

Because of the statistical independence issue discussed in 
\cref{ssec:filter}, we are unable to use the RAS loss in fully unsupervised settings, however, we demonstrate that our proposed method can obtain good speech separation performance with very limited supervised data. We study the impact of the RAS loss in three different data configurations. The WHAMR! training set consists of \num{20000} labeled mixtures. In the first two configurations, we restrict the number of supervised training examples (labeled data) to \num{500} and \num{1000} samples. These samples are used to train supervised baseline models denoted as ``Sup.~(500)'' and ``Sup.~(1000),'' respectively. Along with these supervised examples, the remaining two-channel mixtures (\num{19500} and \num{19000} samples, respectively) are used for unsupervised fine-tuning using our proposed RAS approach. In a third configuration, we use the full training set of \num{20000} labeled mixtures to train a supervised topline model, denoted as ``Sup.~(full).'' This represents the best possible speech separation for our model configuration and hyperparameters.

\begin{table}[t]
\centering
    \sisetup{
    detect-weight, %
    mode=text, %
    tight-spacing=true,
    round-mode=places,
    round-precision=1,
    table-format=1.1
    }
\caption{Separation and objective speech quality metrics for semi-supervised reverberant speech separation. Metrics are shown for the raw output of the network and after applying a Wiener filter to fit the left-channel mixture $m_L$, in the format ``raw output $|$ after Wiener filter''. SDR and SI-SDR are in \si{dB}. Numbers in parentheses denote the amount of supervised samples available for training. For RAS, all other samples are used as unlabeled data. %
}\label{tab:results} \vspace{-.3cm}
\setlength{\tabcolsep}{4pt}
\begin{tabular}{l
    *{2}{S@{\,\( | \)\,}S}
    *{2}{S[round-precision=2,table-format=1.2]@{\,\( | \)\,}S[round-precision=2,table-format=1.2]}}
\toprule
     Model              & \multicolumn{2}{c}{SDR}   & \multicolumn{2}{c}{SI-SDR}  & \multicolumn{2}{c}{PESQ}  & \multicolumn{2}{c}{STOI} \\
\midrule
Sup.~(500) baseline   & 2.81  & 1.97     & 2.19 & 2.98     & 1.68  & 1.65 & 0.69    & 0.68  \\
Sup.~(500) + RAS  
& \bfseries 5.1   & 4.6              & 4.1 & \bfseries 5.3   & 1.65   & \bfseries 1.69 & \bfseries 0.71     & \bfseries 0.71   \\
\midrule
Sup.~(1000) baseline       
& 3.91     & 2.64     & 3.43  & 3.31        & \bfseries 1.91  & 1.85   & 0.73  & 0.72 \\
Sup.~(1000) + RAS 
& \bfseries 5.92    & 5.37    & 4.92    & \bfseries  5.94     & 1.73   & 1.79   & \bfseries 0.74      & \bfseries 0.74  \\
\midrule
Sup.~(full) topline     
& 7.50       & 6.40           & 7.55   & 6.98    & 2.36   & 2.29  & 0.83        & 0.82  \\
\bottomrule
\end{tabular}\vspace{-.3cm}
\end{table}

The speech separation model consists of a four-layer bi-di\-rec\-tion\-al LSTM with \num{600} hidden units in each layer. We use dropout with a probability of \num{0.3} in each layer.
The BLSTM predicts a phase-sensitive approximation (PSA) mask~\cite{erdogan2015mask} for each source.
The input to the network is the log of the STFT magnitude of the observed mixture computed with a window size of \SI{32}{ms} and a hop size of \SI{8}{ms}.
In the RAS case, we first run a pre-training stage on the supervised data, and then a second training stage using the semi-supervised objective combining $\mathcal{L}_{\operatorname{SI-SDR}}$ on the labeled data with $\mathcal{L}_{\operatorname{RAS}}$ on the unlabeled data. Each stage of training is done for a maximum of \num{100} epochs.
The Adam optimizer is used with an initial learning rate of \num{0.0001}.
The learning rate is decayed by a factor of \num{0.5} whenever the validation loss stagnates for more than \num{5} epochs.

We evaluate performance using BSSEval SDR~\cite{vincent2006bss} using the dry ground truth source signals as reference. We also compute SI-SDR~\cite{leroux2019sdr}, PESQ~\cite{rix2001pesq}, and STOI~\cite{H.Taal2011} using the ground-truth reverberant signal as reference. 

\vspace{-.2cm}
\subsection{Results}
\vspace{-.1cm}

\Cref{tab:results} shows the improvement obtained by applying RAS on models trained with limited supervised data. We present metrics on the raw output of the network, as well as after applying a Wiener filter that fits the raw output to the \emph{left-channel} mixture $m_L$. This is different from what is done at training time (where we have access to the right-channel mixture), so that the inference procedure remains single-channel, only considering the left-channel input.
In the supervised case, computing metrics after the Wiener filter causes a mismatch between training and test conditions and can thus be ignored, but we include them in Table~\ref{tab:results} for completeness.
All metrics are computed using the left\hyp{}channel signals (either dry or reverberant depending on the metric) as reference.
RAS significantly improves separation quality, with a \num{2}~-~\SI{3}{\decibel} improvement in SDR and SI-SDR. Wiener filtering to fit the left-channel mixture results in a decrease of SDR and an increase in SI-SDR. This is reasonable as SDR compares to the dry source signal and compensates for the channel, so reintroducing a channel effect via Wiener filtering is likely detrimental; on the other hand, SI-SDR compares to the reverberant source and does not compensate for channel mismatch, so fitting the raw output to the mixture via filtering helps match the channel and increase the score. For the speech quality metrics,
STOI shows small improvements in both data settings, while PESQ decreases slightly compared to the supervised model trained with 1000 samples.
We noticed that RAS helps suppress the interfering speaker, which leads to the observed gain in time-domain SDR metrics, but may at times also over-suppress the interfering speaker which is heavily penalized by PESQ. Wiener filtering to fit the mixture mitigates this issue and reduces the gap.

In \cref{tab:anechoic}, we show that RAS can be used for domain adaptation from anechoic to reverberant data. In these experiments, we only use the supervised anechoic data (which in practice is typically easier to obtain than labeled reverberant mixtures) from WHAMR! and finetune on reverberant mixtures using RAS. When comparing the ``Sup.~(full) topline'' rows between \cref{tab:results} and \cref{tab:anechoic}, as expected, we observe better performance on reverberant test data when using reverberant training data (\cref{tab:results}). However, when using limited amounts of supervised data, we see that models trained on anechoic data (\cref{tab:anechoic}) actually perform better than those trained on reverberant data (\cref{tab:results}), possibly because when learning to separate with limited training data, anechoic mixtures are easier to learn a model of speech from few examples. The trends between RAS and the supervised models are similar between \cref{tab:results} and \cref{tab:anechoic}.

\begin{table}[t]
\centering
    \sisetup{
    detect-weight, %
    mode=text, %
    tight-spacing=true,
    round-mode=places,
    round-precision=1,
    table-format=1.1
    }
\caption{Reverberant speech separation results when using anechoic supervised samples and reverberant unsupervised samples for training. Metrics are on raw output $|$ after Wiener filter to fit $m_L$. SDR and SI-SDR are in dB.}\label{tab:anechoic} \vspace{-.3cm}
\setlength{\tabcolsep}{4pt}
\begin{tabular}{l
    *{2}{S@{\,\( | \)\,}S}
    *{2}{S[round-precision=2,table-format=1.2]@{\,\( | \)\,}S[round-precision=2,table-format=1.2]}}
\toprule
     Model              & \multicolumn{2}{c}{SDR}   & \multicolumn{2}{c}{SI-SDR}  & \multicolumn{2}{c}{PESQ}  & \multicolumn{2}{c}{STOI} \\
\midrule
Sup.~(500) baseline       & 3.2    & 2.1         & 2.7      & 2.7  & \bfseries 1.86     & 1.80  & \bfseries 0.72       & 0.71  \\
Sup.~(500) + RAS  & \bfseries 5.3    & 4.9         & 4.3      & \bfseries 5.5  & 1.68     &  1.72  & \bfseries 0.72     &  \bfseries 0.72   \\
\midrule
Sup.~(1000) baseline      & 4.0     & 2.9          & 3.5     & 3.5    &  \bfseries 1.83    & 1.79    & 0.73         & 0.72 \\
Sup.~(1000) + RAS &  \bfseries 5.8    & 5.1           & 4.7   &  \bfseries 5.7     & 1.74    & 1.78  & 0.73       & \bfseries 0.74  \\
\midrule
Sup.~(full) topline& 6.9     & 5.80           & 6.6      & 6.0   & 2.23    & 2.21   & 0.81      & 0.80  \\
\bottomrule
\end{tabular}%
\end{table}

\begin{table}[t]
\centering
    \sisetup{
    detect-weight, %
    mode=text, %
    tight-spacing=true,
    round-mode=places,
    round-precision=1,
    table-format=1.1
    }
\caption{\label{tab:ablation} Ablation experiments for RAS loss function. Metrics are on the raw output of the network using 1000 supervised examples.}\vspace{-.3cm}
\begin{tabular}{l
    *{2}{S}*{2}{S[round-precision=2,table-format=1.2]}}
\toprule
Ablation  & {SDR} & {SI-SDR} & {PESQ} & {STOI} \\ \midrule
\textbf{Filter:} indep. $\Rightarrow$ joint & 3.1 & 1.5    & 1.72 & 0.70 \\
\textbf{Loss:} SI-SDR $\Rightarrow$ SNR     & 5.5 & 4.5    & 1.67 & 0.71 \\
\textbf{Threshold:} 10 $\Rightarrow$ $\infty$  & 5.8 & 4.8    & \bfseries 1.73 & \bfseries  0.74     \\ \midrule
Proposed & \bfseries 5.9 & \bfseries 4.9    & \bfseries 1.73 & \bfseries 0.74 \\
\bottomrule
\end{tabular}
\vspace{-.2cm}
\end{table}

\Cref{tab:ablation} shows ablation results for our proposed method.
If we replace the independent Wiener filter estimation for the two estimated sources with joint filter estimation, we observe a drop in all metrics.
This is consistent with the discussion in \cref{ssec:filter} regarding the difficulty of obtaining consistent gradient directions for joint filter estimation.
Replacing the SI-SDR loss in \cref{eq:ras} with SNR (i.e., removing scale invariance) also reduces separation quality.
While the Wiener filter could be expected to absorb any scale ambiguities, in practice the added flexibility from the scale-invariant loss proves valuable.
Filtering out mixtures used for fine-tuning (using a threshold on SDR($m_R(t)$, $\hat{m}_R(t)$) as described in \cref{ssec:rasObjective}) leads to a slight boost in SDR, but it is also beneficial as it reduces training time by not including unsupervised examples for which RAS separation is not expected to work well.

\vspace{-.15cm}
\section{Conclusion}
\vspace{-.05cm}

We proposed reverberation as separation (RAS), a semi-supervised approach for reverberant speech separation. We have shown that RAS significantly improves speech separation when there is limited labeled data, and that it is a viable approach for domain adaptation. Future work includes improving the design of the filtering procedure to introduce a stronger inductive bias that further promotes separation of the source estimates, and exploring the possibility of a fully unsupervised version of RAS. 

\vfill\pagebreak
\balance 
\bibliographystyle{IEEEtran}
\bibliography{refs}

\end{document}